LOO and WAIC as Model Selection Methods for Polytomous Items

Yong Luo

National Center for Assessment in Saudi Arabia

## Abstract

Watanabe-Akaike information criterion (WAIC; Watanabe, 2010) and leave-one-out cross validation (LOO) are two fully Bayesian model selection methods that have been shown to perform better than other traditional information-criterion based model selection methods such as AIC, BIC, and DIC in the context of dichotomous IRT model selection. In this paper, we investigated whether such superior performances of WAIC and LOO can be generalized to scenarios of polytomous IRT model selection. Specifically, we conducted a simulation study to compare the statistical power rates of WAIC and LOO with those of AIC, BIC, AICc, SABIC, and DIC in selecting the optimal model among a group of polytomous IRT ones. We also used a real data set to demonstrate the use of LOO and WAIC for polytomous IRT model selection. The findings suggest that while all seven methods have excellent statistical power (greater than 0.93) to identify the true polytomous IRT model, WAIC and LOO seem to have slightly lower statistical power than DIC, the performance of which is marginally inferior to those of the other four frequentist methods.

## Introduction

Item response theory (IRT; Lord, 1980), a family of mathematical models relating the probability of correct item responses to item characteristics and examinees' abilities, is currently the dominant measurement framework in large-scale testing. Comparing to the classical test theory (CTT), IRT boasts theoretical advantages such as the parameter invariance property (Hambleton & Swaminathan, 1985). Such advantages, however, can materialize only when the IRT model fits the data adequately (Cohen & Cho, 2016). In other words, before the benefits offered by IRT can be reaped, model evaluating endeavors that investigate the alignment between the IRT model and data need to be conducted.

IRT model evaluation consists of model comparison and model fit check, two complementary procedures that assess absolute and relative fit of the proposed IRT model respectively. As IRT can be estimated with both frequentist (e.g., Bock & Aitkin, 1981) and Bayesian methods (e.g., Patz & Junker, 1992a, 1992b), model fit check and model comparison procedures can be either frequentist or Bayesian. To date, both IRT model fit check and IRT model comparison remain two actively-researched lines of research in the psychometric literature (e.g., Chalmers & Ng, 2017; Li, Jiao, & Xie, 2017; Luo & Al-Harbi, 2017). In this paper, we focus our discussion on model comparison techniques, both frequentist and Bayesian, for IRT models. Readers are referred to Glas (2016) and Sinharay (2016) for comprehensive and relatively recent reviews of frequentist and Bayesian IRT model fit analysis.

 For model comparison purposes, information criteria based methods such as Akaike's information criterion (AIC; Akaike, 1973, 1974), Bayesian information criterion (BIC; Shwarz, 1978), and deviance information criterion (DIC; Spiegelhalter, Best, Carlin, & van der Linde, 2002) have been routinely used to choose the best fitting model among a group of candidate ones. Despite their popularity, Vehtari, Gelman, and Gabry (2017) pointed out that these methods are not fully Bayesian and recommended the use of two emerging model selection methods, namely leave-one-out cross-validation (LOO) and widely available information criterion (WAIC; Watanabe, 2010), due to their fully Bayesian nature.

Luo and Al-Harbi (2017) investigated whether such a fully Bayesian nature of LOO and WAIC translated into superior performances in the context of dichotomous IRT model selection. They found that LOO and WAIC had higher statistical power than likelihood ratio test (LRT), AIC, BIC, and DIC, especially when the generating model was the three-parameter logistic (3PL) model. However, whether the superior performances of LOO and WAIC in the case of dichotomous IRT model selection can be generalized to the polytomous case remains unknown. As polytomous items have become a ubiquitous presence in educational and psychological testing (Ostini & Nering, 2010) due to their provision of richer information than dichotomous ones (e.g., Cohen, 1983; Samejima, 1975, 1979), it is important that a proper model selection method should be used to choose a proper polytomous IRT model for the analysis of data based on responses to polytomous items.

The purpose of this study is to investigate the statistical power of LOO and WAIC as model selection methods in choosing the optimal polytomous IRT model, in comparison with other five model selection methods. These five model selection methods include AIC, BIC, DIC, AIC corrected for bias (AICc; Sugiura, 1978), and sample-size-adjusted BIC (SABIC; Sclove, 1987). The rest of this article is organized as follows. First, we describe each of the seven model selection methods (AIC, AICc, BIC, SABIC, DIC, LOO, and WAIC) adopted in the current study and explain why a method is considered non-Bayesian, partially Bayesian, or fully Bayesian. Second, we provide a literature review of simulation studies that investigate performances of various model selection methods in the IRT context, with an emphasis on Luo and Al-Harbi (2017) study. Third, we present a simulation study conducted to compare the statistical power of LOO and WAIC in selecting the correct model with the other five methods. The fourth section is an illustration with a real dataset whether the seven methods produce inconsistent results regarding the choice of the best-fitting model. We conclude this article with discussions and practical suggestions regarding the use of model selection methods with polytomous data.

## Model Comparison Methods

AIC, BIC, AICc and SABIC are all information criterion indices based on maximum likelihood estimation (MLE) that can be expressed as the sum of a deviance term and a penalty term. Specifically, they are computed as

$$AIC = -2\log p(y \,|\, \hat{\theta}_{mle}) + 2k \,, \tag{1}$$

$$BIC = -2\log p(y \,|\, \hat{\theta}_{mle}) + k * \ln(N) \,, \tag{2}$$

$$AICc = -2\log p(y \,|\, \hat{\theta}_{mle}) + 2k \frac{N}{N-k-1} \,, \tag{3}$$

and

$$SABIC = -2\log p(y \,|\, \hat{\theta}_{mle}) + k * \ln(\frac{N+2}{24}) \,. \tag{4}$$

As can be seen, the four model comparison indices share the same deviance term $-2\log p(y \,|\, \hat{\theta}_{mle})$, in which $\hat{\theta}_{mle}$ is the MLE-based point estimate and $\log p(y \,|\, \hat{\theta}_{mle})$ is the log

likelihood of data based on $\hat{\theta}_{mle}$. For the penalty term, AIC uses $2k$ and BIC uses $k * \ln(N)$, with $k$ being the number of parameters and $N$ the sample size. AICc is a variation of AIC that is corrected for bias inherent in AIC when the ratio of sample size N and number of parameters k is small. SABIC is a variation of BIC that penalizes model complexity (as expressed by number of parameters $k$) less harshly than BIC. It should be noted that AICc and SABIC have been rarely used in simulation studies on IRT model selection with one exception (Choi, Paek, & Cho, 2017), where the performances of AIC, BIC, AICc, and SABIC as model selection methods were investigated in the case of mixture Rasch models. As the above four model comparison methods compute their deviance term with MLE point estimate and their penalty term with only sample size and number of parameters, they are typically considered non-Bayesian.

DIC and WAIC are Bayesian information criterion indices that can also be expressed as the sum of a deviance term and a penalty term. Specifically, DIC is computed as

$$DIC = -2\log p(y \mid \hat{\theta}_{EAP}) + 2 p_{DIC}, \tag{5}$$

and

$$p_{DIC} = 2(\log p(y \mid \hat{\theta}_{EAP}) - E_{post}(\log p(y \mid \theta))). \tag{6}$$

As can be seen in equation 5, DIC uses $\hat{\theta}_{EAP}$, a point estimate based on the posterior mean estimate, to compute the log likelihood of data $\log p(y \mid \hat{\theta}_{EAP})$. The penalty term is computed in equation 6, where the second term in the parenthesis is the posterior mean of log likelihood of data and its computation involves the use of the whole posterior distribution of $\theta$. As only the penalty term uses the posterior distribution, DIC is considered partially Bayesian.

The deviance term used in the computation of WAIC requires log pointwise predictive density (LPPD), which is computed as

$$LPPD = \sum_{i=1}^{n} \log \int p(y_i \mid \theta) p_{post}(\theta) d\theta . \tag{7}$$

As the computation of LPPD uses the whole posterior distribution $p_{post}(\theta)$, LPPD can be viewed as a fully Bayesian analog of $\log p(y \mid \hat{\theta}_{mle})$ in the computation of AIC and BIC and $\log p(y \mid \hat{\theta}_{EAP})$ in the computation of DIC. Similar to LPPD, the penalty term of WAIC is fully Bayesian and can be expressed as

$$p_{WAIC} = \sum_{i=1}^{n} \mathrm{var}_{post}(\log p(y_i \mid \theta)) , \tag{8}$$

where the penalty term $p_{WAIC}$ is "the variance of individual terms in the log predictive density summed over the $n$ data points" (Gelman, Carlin, Stern, & Rubin, 2014, p. 173). WAIC is computed as

$$WAIC = -2LPPD + 2p_{WAIC}. \tag{9}$$

LOO differs from the aforementioned information criterion based indices in that its computation requires no penalty term. Specifically, LOO is computed as

$$LOO = -2LPPD_{loo} = -2\sum_{i=1}^{n} \log \int p(y_i \mid \theta) p_{post(-i)}(\theta) d\theta , \tag{10}$$

where $p_{post(-i)}(\theta)$ is the posterior distribution based on the data minus data point $i$. Unlike LPPD that uses data point $i$ is for both the computation of posterior distribution and the prediction, here LPPD$_{loo}$ only uses it for prediction, and hence there is no need for a penalty term to correct the potential bias introduced by using data twice.

### Previous Simulation Studies on IRT Model Selection

While there are many studies applying model selection methods to choose the best-fitting IRT model (e.g., May, 2006; Hickendorff, Heiser, van Putten, & Verhelst, 2009; Revuelta, 2008; Rijmen & De Boeck, 2002; Yao & Schwarz, 2006), relatively few use simulation studies to systematically investigate their performances. Among those simulation studies, some focus on unidimensional IRT (UIRT) model selection (Kang & Cohen, 2007; Kang, Cohen, & Sung, 2009; Luo & Al-Harbi, 2017; Whittaker, Change, & Dodd, 2012), some on multidimensional IRT (MIRT; Reckase, 2009) model selection (Li, Bolt, & Fu, 2006; Revuelta & Ximénez, 2017; Zhu & Stone, 2012), and some on mixture IRT (e.g., Rost, 1990) model selection (e.g., Choi, Paek, & Cho, 2017; Li, Cohen, Kim, & Cho, 2009; Preinerstorfer & Formann, 2012)

In the case of UIRT model selection. Kang and Cohen (2007) investigated the performances of likelihood ratio test (LRT), AIC, BIC, DIC, and the cross-validation log-likelihood (CVLL; O'Hagan, 1995) as model selection methods for dichotomous IRT models. They found that CVLL had the best overall performance, and the five methods sometimes disagreed with each other. In Kang, Cohen, and Sung (2009) study, they compared the performances of AIC, BIC, DIC, and CVLL in choosing the correct polytomous IRT model among the graded response model (GRM; Samejima, 1969), the rating scale model (RSM;

Andrich, 1978), the partial credit model (PSM; Masters, 1982), and the generalized partial credit

model (GPCM; Muraki, 1992). They found that AIC and BIC performed better than CVLL and

DIC, a finding which is in contrary to what was found in Kang and Cohen (2007) study.

Whittaker, Chang, and Dodd (2012) compared the performances LRT, AIC, BIC, AICc, Hannon

and Quinn's information criterion (HQIC; Hannon & Quinn, 1979), and consistent AIC (CAIC;

Bozdogan, 1987) as IRT model selection methods with mixed-format data. They found that no

method performed consistently well, and which method to choose depended on conditions such

as sample size and the ratio of dichotomous and polytomous items. Luo and Al-Harbi (2017)

compared the performances of LOO and WAIC with LRT, AIC, BIC, and DIC as model selection methods for dichotomous IRT models. Similar to Kang and Cohen (2007) study, they manipulated sample size (500, 1000), test length (20, 40), ability distribution with different means (-1, 0, 1), and generating IRT models (1PL, 2PL, 3PL), which resulted in a fully-crossed simulation design with 36 simulation conditions. They found that for LOO and WAIC the average power to identify the correct dichotomous IRT model was 0.98, DIC 0.93, LRT 0.88,

AIC 0.85, and BIC 0.67. In addition, they found that when the generating model was the 3PL and the ability distribution was $N(1,1)$, LRT, AIC, BIC, and DIC performed poorly with average power all less than 0.5, while the power of LOO and WAIC was 0.89 and 0.94, respectively.

They concluded that the fully Bayesian nature of LOO and WAIC does result in superior performances in the context of dichotomous IRT model selection.

 In terms of MIRT model selection, Li, Bolt, & Fu (2006) investigated the performances

of DIC, PsBF, and PPMC to choose the correct model among a group of testlet models. They

found that PsBF and PPMC performed equally well, and DIC performed noticeably worse. Zhu

& Stone (2012) compared the performances of DIC, PPMC, and conditional predictive ordinate (CPO) in selecting the correct model among a group of unidimensional and multidimensional models based on the GRM, which include the one-parameter GRM (Muraki, 1990), multidimensional GRM with simple and complex structures, and the graded response testlet model. They found that all three methods performed equally well, and CPR and PPMC were more versatile than DIC in that they provided fit information at the item level. Revuelta and Ximénez (2017) compared the performances of standardized generalized dimensionality discrepancy measure (SGDDM; Levy, Xu, Yel, & Svetina, 2015), DIC, WAIC, and LOO in assessing dimensionality for the multidimensional nominal response model (MNRM). They found that the PPMC-based SGDDM performed considerably better than the other three methods, among which WAIC and LOO outperformed DIC, and they concluded that for MNRM, SGDDM should be used.

In terms of mixture IRT model selection, Li, Cohen, Kim, and Cho (2009) compared the performances of AIC, BIC, DIC, posterior predictive model checks (PPMC; Gelman, Meng, & Stern, 1996), and the pseudo-Bayes factor (PsBF; Gerisser & Eddy, 1979) in the context of mixture IRT model selection. They found that BIC and PsBF performed better than AIC and PPMC, which were more likely to choose more complex models, and DIC was the least effective method among all. Preinerstorfer and Formann (2012) investigated the performances of AIC and BIC in selecting mixture Rasch models that were estimated with conditional maximum likelihood estimation, and they found that BIC performed better than AIC. Choi, Paek, and Cho (2017) manipulated class-distinction features in a two-class mixture Rasch model and compared the performances of AIC, BIC, AICc, and SABIC under different manipulated conditions. In contrary to the previous findings that BIC consistently performed better than AIC, they found that these four methods performed differentially with different class-distinction features. In addition, they found that AICc and SABIC performed better than or equally with AIC and BIC, respectively.

## Methods

Simulation Design

Following the simulation design in Kang, Cohen, and Sung (2009) study, we manipulated sample size (SS; 500, 1000), test length (TL; 10, 20), number of response categories (NC; 3, 5), and the generating model (GM; GRM, RSM, PCM, GPCM), which results in a fully crossed simulation design with 2*2*2*4=32 conditions. Within each condition 100 datasets were generated based on a data generation procedure described later in this section.

The outcome variable of the current simulation study is statistical power of each of the seven model selection methods. To compute the statistical power of a given method, we record within each simulation condition how many times the true model is selected based on that method and divide that number by 100.

Four Polytomous IRT Models

Common polytomous IRT models for ordinal data include GRM, RSM, PCM, and GPCM. For norminal data such as multiple-choice item response data in educational testing, the nominal response model (NRM; Bock, 1972) is widely used, although it should be noted that there are other options such as the multiple-choice model (Thissen & Steinberg, 1984), the nested logit model (Suh & Bolt, 2010), and the sequential IRT model (Deng & Bolt, 2016). In this paper we focus on the four polytomous IRT models for ordinal data and provide a brief description in the following.

These four IRT models can be divided into the difference model and the divide-by-total model (Thissen & Steinberg, 1986). As the only difference model, GRM first models the probability of responding below a certain category vs above that category; the probability of responding at that category is then computed as the difference of the two probabilities. The mathematical equation for GRM is given as

$$p_{ij}(u_{ij} = k \mid \theta_i, a_j, b_{jk}) = \frac{1}{1 + \exp(-a_j(\theta_i - b_{jk}))} - \frac{1}{1 + \exp(-a_j(\theta_i - b_{jk+1}))} , \tag{11}$$

where $p_{ij}$ is the probability of responding in a category $k$ or higher, $u_{ij}$ is the response of examinee $i$ to item $j$, $\theta_i$ is the latent proficiency of examinee $i$, and $a_j$ and $b_{jk}$ are the item discrimination and the category difficulty of item $j$.

GPCM, PCM, and RSM are all divide-by-total models. GPCM, as the name suggests, is a generalized case of the partial credit model (PCM; Masters, 1982). The probability of responding in category $k$ based on GPCM is

$$p_{ij}(u_{ij} = k \mid \theta_i, a_j, \delta_{jk}) = \frac{\exp[\sum_{h=1}^{k_j} a_j(\theta_i - (\delta_j - \tau_{jh}))]}{\sum_{c=1}^{m_j} \exp[\sum_{h=1}^{c} a_j(\theta_i - (\delta_j - \tau_{jh}))]} , \tag{12}$$

where $\delta_j$ is the item location parameter of item $j$, $\tau_{jh}$ is the step parameter for category h of item $j$, $m_j$ is the number of categories of item $j$, and the other terms remain the same as in equation 10. PCM can be obtained by constraining $a_j$ to be 1 across all items; RSM can be obtained by further holding $\tau_{jh}$ to be constant across items.

Data Generation

Item parameter values used for data generation are listed in Table 1. As can be seen, the first 20 items have three response categories and the last 20 have five categories. Correspondingly, when NC=3, the first 20 items were used for data generation; when NC=5, the last 20 were used. When GM=PCM, the item discrimination parameter $a$ was fixed to one; when GM=RSM, in addition to the constraint imposed in PCM, only the threshold parameters of the first GPCM item with three categories (item 1) and the first GPCM item with five categories (item 21) were used. It should be noted that the item parameter values in Table 1 were deliberately chosen to be the same as in Kang, Cohen, and Sung (2009) to facilitate the comparison between the findings in their study and the current one by eliminating the potential confounding effect of using different generating item parameter values.

Table 1

*Generating Item Parameters*

| Item | GRM | | | | | GPCM | | | | |
|------|------|------|------|------|------|------|------|---------|---------|---------|
| | a | b1 | b2 | b3 | b4 | a | b | $\tau_1$ | $\tau_2$ | $\tau_3$ |
| 1 | 1.19 | -1.21 | 1.77 | | | 1.16 | -0.42 | 1.26 | | |
| 2 | 0.96 | -1.32 | 1.22 | | | 0.51 | -0.24 | 0.66 | | |
| 3 | 1.52 | -0.36 | 1.84 | | | 1.43 | 0.61 | 1.47 | | |
| 4 | 2.48 | -0.62 | 1.82 | | | 2.25 | -0.37 | 0.74 | | |
| 5 | 0.58 | -1.49 | 0.22 | | | 0.71 | 0.16 | 1.23 | | |
| 6 | 1.13 | -2.96 | 0.59 | | | 1.54 | 0.6 | 0.76 | | |
| 7 | 1.63 | 0.24 | 2.21 | | | 1.87 | 0.11 | 0.52 | | |
| 8 | 0.82 | -2.41 | 0.81 | | | 0.45 | -0.4 | 0.65 | | |
| 9 | 1.97 | -2.38 | 0.46 | | | 0.49 | -0.38 | 1.57 | | |
| 10 | 1.21 | -2.08 | 1.17 | | | 1.33 | 0.15 | 0.72 | | |
| 11 | 1.1 | -1.78 | 1.04 | | | 0.82 | -0.19 | 0.91 | | |
| 12 | 0.8 | 0.68 | 2.43 | | | 1.41 | -0.03 | 0.67 | | |
| 13 | 2.02 | -2.1 | 0.93 | | | 1.5 | 0.36 | 1.18 | | |
| 14 | 1.85 | -0.21 | 1.42 | | | 1.43 | 0.35 | 0.52 | | |
| 15 | 1.48 | -1 | 1.69 | | | 1.91 | -0.29 | 1.03 | | |
| 16 | 1.4 | -1.97 | 0.15 | | | 1.4 | -0.34 | 0.97 | | |
| 17 | 2.47 | -1.51 | 1.91 | | | 1.81 | 0.16 | 0.79 | | |
| 18 | 0.93 | -1.35 | 0.85 | | | 0.55 | -0.25 | 0.98 | | |
| 19 | 1.24 | -1.14 | 2.25 | | | 0.99 | 0.21 | 0.37 | | |
| 20 | 1.65 | -1.1 | 1.31 | | | 0.92 | 0.19 | 1.27 | | |
| 21 | 1.19 | -1.59 | -0.83 | 1.25 | 2.28 | 1.16 | -0.42 | 2.56 | -0.04 | -1.67 |
| 22 | 0.96 | -2.35 | -0.29 | 0.6 | 1.84 | 0.51 | -0.24 | 0.88 | 0.45 | -1.67 |
| 23 | 1.52 | -0.67 | -0.06 | 1.28 | 2.39 | 1.43 | 0.61 | 3.05 | -0.1 | -0.95 |
| 24 | 2.48 | -1.2 | -0.04 | 1.22 | 2.42 | 2.25 | -0.37 | -0.41 | 1.88 | 0 |
| 25 | 0.58 | -1.84 | -1.13 | -0.17 | 0.62 | 0.71 | 0.16 | 2.35 | 0.11 | -0.67 |
| 26 | 1.13 | -3.68 | -2.23 | -0.3 | 1.48 | 1.54 | 0.6 | 1.45 | 0.08 | -0.26 |
| 27 | 1.63 | -0.58 | 1.06 | 1.81 | 2.62 | 1.87 | 0.11 | 1.27 | -0.24 | 0.5 |
| 28 | 0.82 | -3.83 | -0.98 | 0.49 | 1.12 | 0.45 | -0.4 | 1.9 | -0.6 | -0.28 |
| 29 | 1.97 | -3.51 | -1.26 | 0.13 | 0.79 | 0.49 | -0.38 | 3.17 | -0.04 | -2.08 |
| 30 | 1.21 | -2.51 | -1.65 | 0.72 | 1.62 | 1.33 | 0.15 | 1.59 | -0.15 | -0.34 |
| 31 | 1.1 | -2.15 | -1.4 | 0.59 | 1.48 | 0.82 | -0.19 | 2.2 | -0.38 | -1.2 |
| 32 | 0.8 | 0.21 | 1.14 | 2.04 | 2.81 | 1.41 | -0.03 | 0.73 | 0.6 | -0.74 |
| 33 | 2.02 | -3.07 | -1.13 | 0.33 | 1.52 | 1.5 | 0.36 | 1.23 | 1.12 | 0.38 |
| 34 | 1.85 | -0.64 | 0.22 | 1 | 1.83 | 1.43 | 0.35 | 0.03 | 1.02 | 0.28 |
| 35 | 1.48 | -1.97 | -0.03 | 0.96 | 2.41 | 1.91 | -0.29 | 0.49 | 1.56 | -1.36 |
| 36 | 1.4 | -2.64 | -1.3 | -0.33 | 0.63 | 1.4 | -0.34 | 1.68 | 0.27 | -0.02 |
| 37 | 2.47 | -2.09 | -0.94 | 1.42 | 2.4 | 1.81 | 0.16 | 1.16 | 0.42 | -1.24 |
| 38 | 0.93 | -1.91 | -0.79 | 0.44 | 1.26 | 0.55 | -0.25 | 2.14 | -0.18 | -1.44 |
| 39 | 1.24 | -1.61 | -0.66 | 1.66 | 2.85 | 0.99 | 0.21 | 1.6 | -0.86 | 0.41 |
| 40 | 1.65 | -2.05 | -0.16 | 0.67 | 1.96 | 0.92 | 0.19 | 1.62 | 0.92 | -0.16 |

The latent ability was generated from a standard normal distribution $N(0,1)$ and the same set of generated latent ability values were used for the same SS. The item parameter values in Table 1 and the generated ability values were plugged in the corresponding GM equation to generate item response data. It is important to note that after each data set was generated, we checked to make sure that no item contained any null category, a situation which would make the maximum likelihood estimation of RSM problematic. If a generated data set contained item(s) with null categories, it was replaced with a new data set that satisfied this requirement.

Estimation and Computation

The R package **mirt** (Chalmers, 2012) was used for maximum likelihood estimation and computation of AIC, BIC, AICc, and SABIC. For MCMC estimation, we used the R package **rstan**, the R interface to the Bayesian software program Stan (Carpenter et al., 2016). Stan implements the no-U-turn sampler (NUTS; Hoffman & Gelman, 2014), which is an improved version of Hamiltonian Monte Carlo (HMC; Neal, 2011) - a powerful and efficient MCMC algorithm that has been shown to work well for IRT models (e.g., Luo & Jiao, 2017). We adopted priors identical to those used by Kang, Cohen, and Sung (2009) for MCMC estimation. Specifically, for the item discrimination parameter in both GRM and GPCM, a lognormal distribution $ln(0,1)$ was assigned as the prior; for the item location parameter in RSM, PCM, and RSM, a standard normal distribution $N(0,1)$ was used as the prior; a normal distribution $N(0,10)$ was assigned as prior for both the category difficulty parameter in GRM and the threshold parameter in the other three models. For model identification purpose, a standard normal distribution N(0,1) was used as the prior for the latent ability regardless of the IRT model; in addition, for GPCM, PCM, and RSM, the sum of the threshold parameters was constrained to zero.

For computation of WAIC and LOO, we used the R package **loo** (Vehtari et al., 2016b) that computes LOO via Pareto smoothed importance sampling (PSLS; Vehtari, Gelman, & Gabry, 2015). Unlike WinBUGS that can be specified to compute DIC, Stan does not have such a feature and we wrote a R program to extract the posterior draws produced by **rstan** and computed DIC based on equations 5 and 6.

Convergence Check

For MCMC estimation, the Gelman and Rubin's convergence diagnostic (Gelman & Rubin, 1992) that computes the potential scale reduction factor (PSRF) was applied to check model convergence. A PSRF value close to one is considered indicative of model convergence and Gelman, Carlin, Stern, and Rubin (2014) recommend to use 1.1 as the threshold value. We found that in **rstan**, the PSRF values for all four models dropped to below 1.05 after 150 iterations and consequently, we ran three parallel chains with 400 iterations each to make sure that model convergence was not an issue. It is worth noting that the efficient HMC algorithm implemented in Stan, as demonstrated by Luo and Jiao (2017), is the reason why usually several hundred iterations are adequate to reach model convergence for complex IRT models such as multidimensional and multilevel ones. In contrast, Kang, Cohen, and Sung (2009) ran 11,00 iterations in WinBUGS for the same set of IRT models.

**Results**

Model selection results are listed in Table 2. Specifically, it provides the statistical power of a model selection method under each simulation condition. For example, the first row in Table 2 indicates that when GM=GPCM, NRC=3, SS=500, and TL=10, the statistical power of AIC, BIC, AICc, and SABIC to identify GPCM as the true model is 0.91, DIC 0.73, LOO 0.88, and WAIC 0.94.

Table 2

*Power Rates of Different Model Selection Methods*

| True Model | Response Categories | Sample Size | Test Length | Model Selection Methods | | | | | | |
|---|---|---|---|---|---|---|---|---|---|---|
| | | | | AIC | AICc | BIC | SABIC | DIC | LOO | WAIC |
| | | | 10 | 0.91 | 0.91 | 0.91 | 0.91 | 0.73 | 0.88 | 0.94 |
| | | 500 | 20 | 1 | 1 | 1 | 1 | 0.96 | 0.98 | 0.99 |
| | 3 | | 10 | 1 | 1 | 1 | 1 | 0.95 | 0.96 | 0.99 |
| | | 1000 | 20 | 1 | 1 | 1 | 1 | 1 | 1 | 1 |
| GPCM | | | 10 | 1 | 1 | 1 | 1 | 0.87 | 1 | 1 |
| | | 500 | 20 | 1 | 1 | 1 | 1 | 0.94 | 1 | 1 |
| | 5 | | 10 | 1 | 1 | 1 | 1 | 1 | 1 | 1 |
| | | 1000 | 20 | 1 | 1 | 1 | 1 | 1 | 1 | 1 |
| | | | 10 | 0.79 | 0.79 | 0.79 | 0.79 | 0.92 | 0.72 | 0.62 |
| | | 500 | 20 | 0.94 | 0.94 | 0.94 | 0.94 | 0.97 | 0.91 | 0.89 |
| | 3 | | 10 | 0.91 | 0.91 | 0.91 | 0.91 | 0.92 | 0.81 | 0.69 |
| | | 1000 | 20 | 1 | 1 | 1 | 1 | 0.98 | 0.98 | 0.98 |
| GRM | | | 10 | 1 | 1 | 1 | 1 | 1 | 0.99 | 0.96 |
| | | 500 | 20 | 1 | 1 | 1 | 1 | 1 | 1 | 1 |
| | 5 | | 10 | 1 | 1 | 1 | 1 | 1 | 1 | 0.98 |
| | | 1000 | 20 | 1 | 1 | 1 | 1 | 1 | 1 | 1 |
| | | | 10 | 0.94 | 0.96 | 0.83 | 1 | 0.93 | 0.89 | 0.86 |
| | | 500 | 20 | 0.99 | 1 | 0.66 | 1 | 0.98 | 1 | 0.98 |
| | 3 | | 10 | 0.96 | 0.96 | 1 | 0.99 | 0.97 | 0.88 | 0.84 |
| | | 1000 | 20 | 0.99 | 0.99 | 1 | 1 | 0.99 | 0.97 | 0.97 |
| PCM | | | 10 | 0.96 | 0.99 | 1 | 1 | 0.99 | 0.86 | 0.83 |
| | | 500 | 20 | 0.99 | 1 | 1 | 1 | 1 | 0.98 | 0.97 |
| | 5 | | 10 | 0.95 | 0.97 | 1 | 1 | 1 | 0.83 | 0.81 |
| | | 1000 | 20 | 1 | 1 | 1 | 1 | 1 | 0.99 | 0.97 |
| | | | 10 | 0.97 | 0.99 | 1 | 1 | 0.95 | 0.88 | 0.88 |
| | | 500 | 20 | 0.99 | 1 | 1 | 1 | 1 | 0.99 | 0.99 |
| | 3 | | 10 | 0.97 | 0.98 | 1 | 1 | 0.67 | 0.78 | 0.79 |
| | | 1000 | 20 | 1 | 1 | 1 | 1 | 0.97 | 0.99 | 0.99 |
| RSM | | | 10 | 1 | 1 | 1 | 1 | 1 | 1 | 1 |
| | | 500 | 20 | 1 | 1 | 1 | 1 | 1 | 1 | 1 |
| | 5 | | 10 | 1 | 1 | 1 | 1 | 1 | 1 | 1 |
| | | 1000 | 20 | 1 | 1 | 1 | 1 | 1 | 1 | 1 |

In Figure 1 a visual presentation of the mean statistical power comparison of the seven methods is presented. As can be seen, all seven model selection methods have statistical power greater than 0.93. One noticeable pattern is that the frequentist-based methods (AIC, BIC, AICc, and SABIC) seem to perform better than the Bayesian ones (DIC, LOO, and WAIC). Among the frequentist-based ones, AICc (power = 0.981) performs slightly better than AIC (power = 0.977), and SABIC (power = 0.986) better than BIC (power = 0.970). Among the three Bayesian methods, DIC (power = 0.959) performs slightly better than LOO (power = 0.946) and WAIC (power = 0.935).

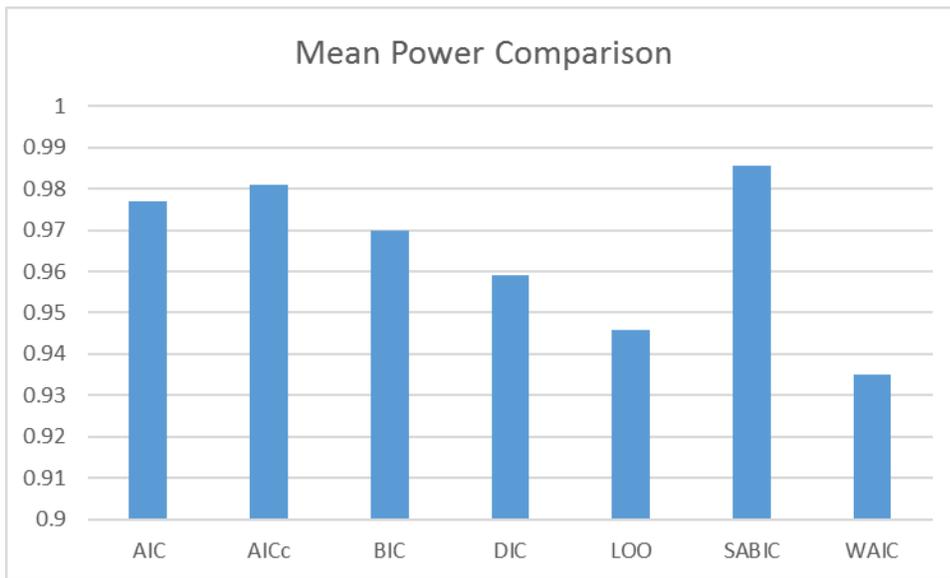

*Figure 1.* Mean power rates of different methods

To gain better insights into how the three manipulated factors (test length, sample size, number of response categories) affect the performance of each method with a given item response generating model, we provide visual presentations of marginal summaries of the performances of the seven methods with bar plots in Figures 2-4. Specifically, we focus on how test length (TL) affects the performance of each model selection method in Figure 2, where the number of times a true model was chosen at a given test length was aggregated over four simulation conditions (the combination of two SSs and two NCs result in 400 simulated datasets). As can be seen, when GM=GPCM and TL=10, LOO and WAIC perform approximately the same as the other four frequentist methods (GPCM is correctly identified close to 400 times), and DIC seems to have a considerably higher probability to choose GRM (more than 40 times) as the true model; the difference in performance between DIC and the other six methods, however, decreases when TL=20. When GM=GRM and TL=10, LOO and WAIC are more likely than the other five methods to choose GPCM as the true model (approximately 50 times for LOO and 70 times for WAIC); when TL=20, the tendency of LOO and WAIC to choose GPCM decreases noticeably. When GM=PCM and TL=10, LOO and WAIC are considerably more likely than the other methods to select GPCM (the more parameterized model) as the true model ((approximately 50 times for LOO and 60 times for WAIC)); when TL=20, the statistical power of LOO and WAIC increases considerably (they identify PCM correctly more than 380 times) and becomes only slightly lower than AIC, AICc, SABIC, and DIC. When GM=RSM and TL=10, LOO and WAIC show a similar pattern of having a higher probability of identifying a more parameterized model (PCM) as the true model; when TL=20, the statistical power of LOO and WAIC increases to almost one. To sum up, Figure 2 shows that when GM=GPCM, LOO and WAIC perform well (better than or equal to other methods) regardless of TL; when GM is not GPCM, LOO and WAIC perform worse than other methods when TL=10, and with the increase of test length (TL=20) LOO and WAIC perform equally well as other methods.

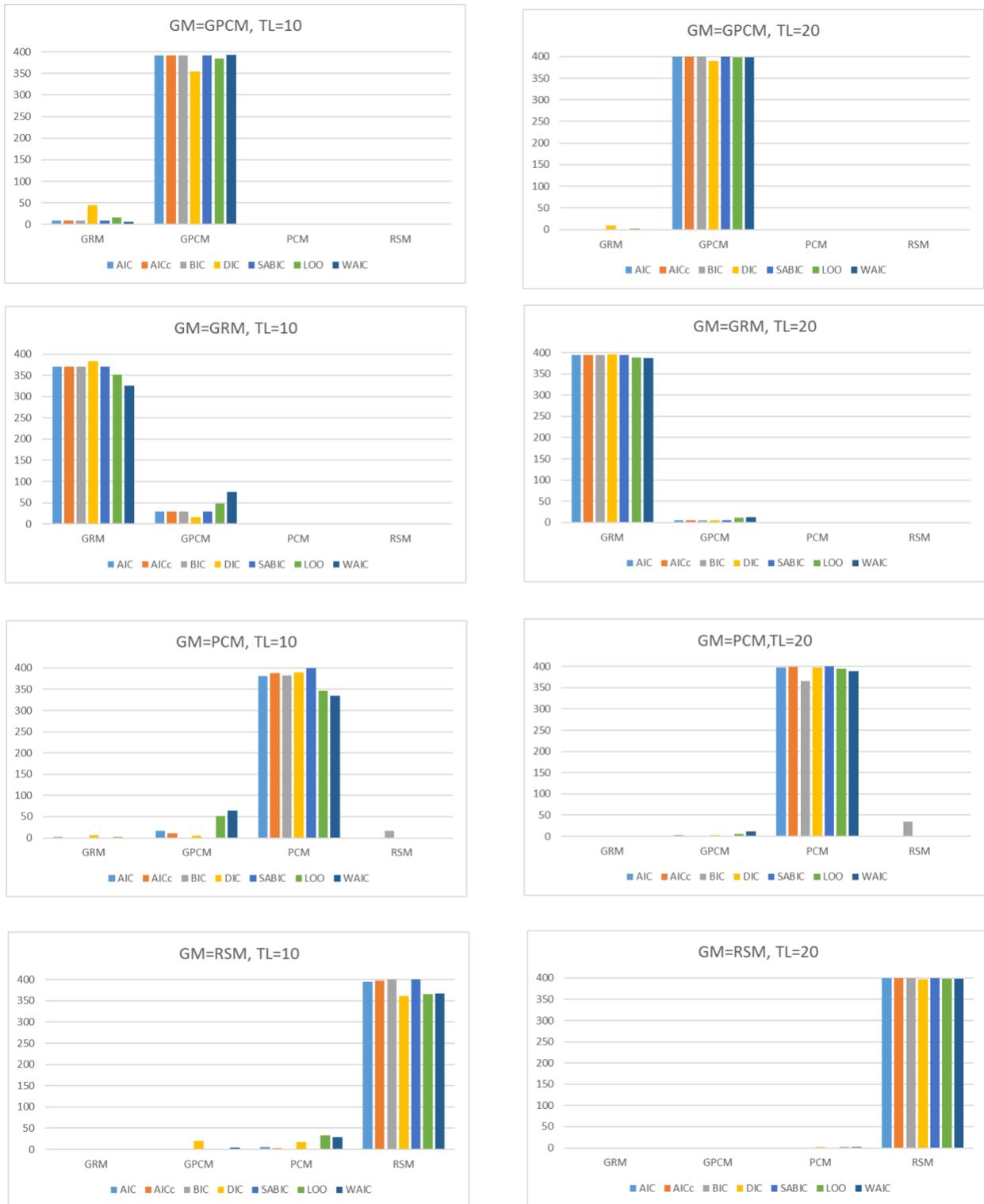

*Figure 2.* Model selection by test length

Figure 3 focuses on how sample size (SS) affects the performance of each model selection method. Specifically, we consider the number of times a true model was chosen with a given sample size aggregated over four simulation conditions (the combination of two TLs and two NCs). As can be seen, when GM=GPCM and SS=500, LOO and WAIC perform approximately the same as the other four frequentist methods, and DIC performs noticeably worse due to its higher probability to choose GRM as the true model; when SS=1000, all seven methods perform similarly and have statistical power close to one. When GM=GRM, LOO and WAIC are slightly more likely than the other five methods to choose GPCM as the true model regardless of SS, and the tendency of LOO and WAIC to choose GPCM decreases with the increase of SS from 500 to 1000. When GM=PCM, LOO and WAIC are considerably more likely than the other methods to select GPCM (the more parameterized model) regardless of SS, and their performance does not improve with the increase of SS from 500 to 1000. When GM=RSM, LOO and WAIC show a similar pattern of having a higher probability of identifying a more parameterized model (PCM) as the true model, and the tendency of LOO and WAIC to choose PCM does not seem to decrease with the increase of SS from 500 to 1000. To sum up, Figure 3 shows that when GM=GPCM, LOO and WAIC perform well (better than or equal to other methods) regardless of TL; when GM=GRM, LOO and WAIC perform worse than other methods regardless of SS, and the performances of LOO and WAIC improve with the increase of SS from 500 to 1000; when GM=PCM or GM=RSM, LOO and WAIC have higher probabilities to choose an incorrect model (more parameterized than GM) than other methods regardless of SS, and the increase of SS from 500 to 1000 does not seem to improve the statistical power of LOO and WAIC.

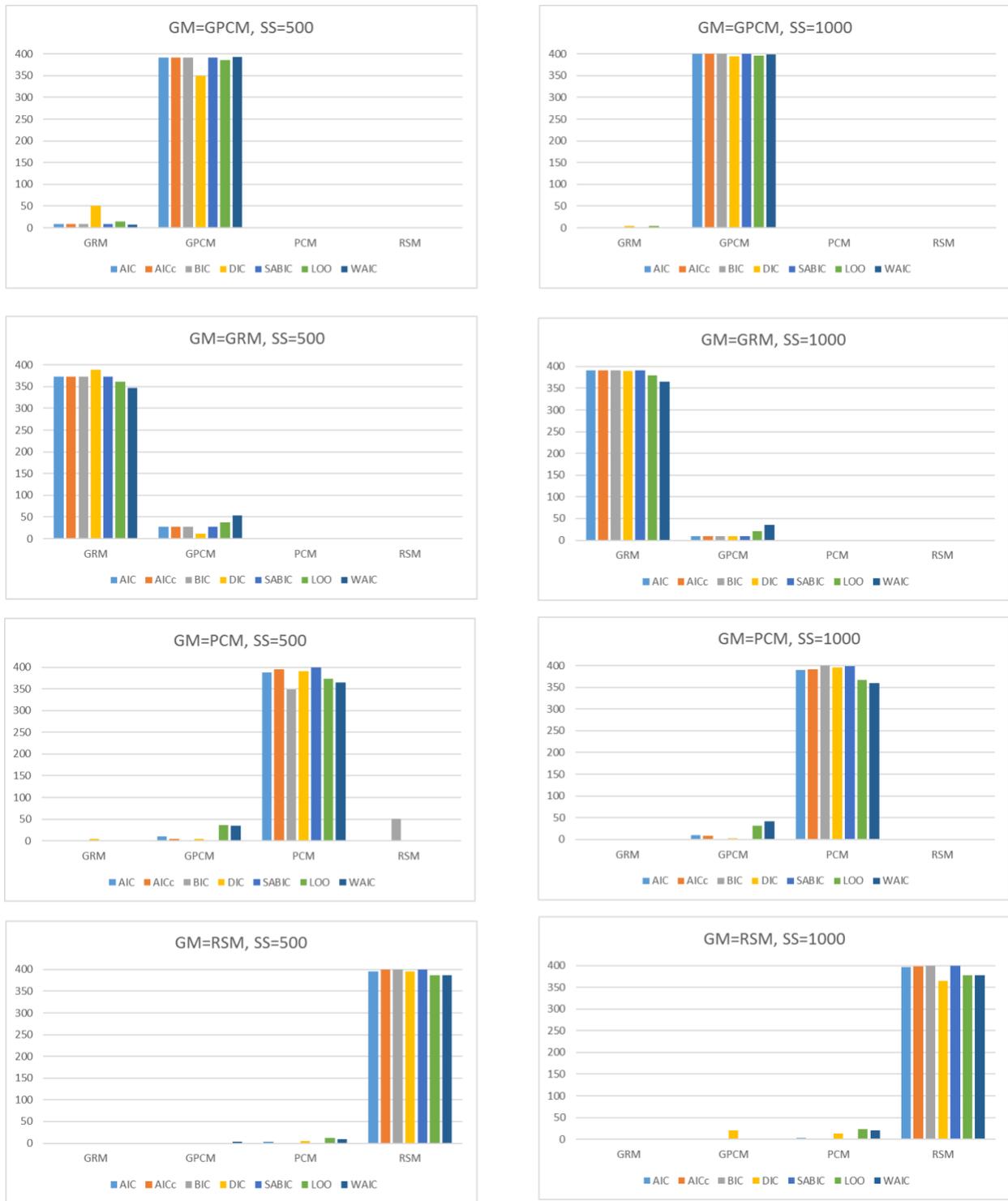

*Figure 3.* Model selection by sample size

Figure 4 focuses on how number of response categories (NC) affects the performance of each model selection method. Specifically, we consider the number of times a true model was chosen with a given number of response categories aggregated over four simulation conditions (the combination of two TLs and two SSs). As can be seen, when GM=GPCM and NC=3, LOO and WAIC perform approximately the same as the other four frequentist methods, and DIC seems to have a considerably higher probability to choose GRM as the true model; the performances of all seven methods improve when NC=5. When GM=GRM and NC=3, LOO and WAIC perform the worst and are more likely than the other five methods to choose GPCM as the true model; when NC=5, however, the tendency of all seven methods to choose GPCM virtually disappears. When GM=PCM and NC=3, LOO and WAIC are considerably more likely than the other methods to select GPCM (the more parameterized model); when NC=5, the performances of LOO and WAIC do not seem to improve, despite that all the other five methods perform better with the increase of NC. When GM=RSM and NC=3, LOO and WAIC have higher probabilities to identify PCM as the true model; when NC=5, the statistical power of all seven methods become one. To sum up, Figure 4 shows that when GM=GPCM, LOO and WAIC perform well (better than or equal to other methods) regardless of NC; when GM=PCM, LOO and WAIC perform consistently worse than other methods regardless of NC; when GM=GRM or GM=RSM, LOO and WAIC perform worse than other methods when NC=3, and when NC=5, LOO and WAIC perform equally well as other methods.

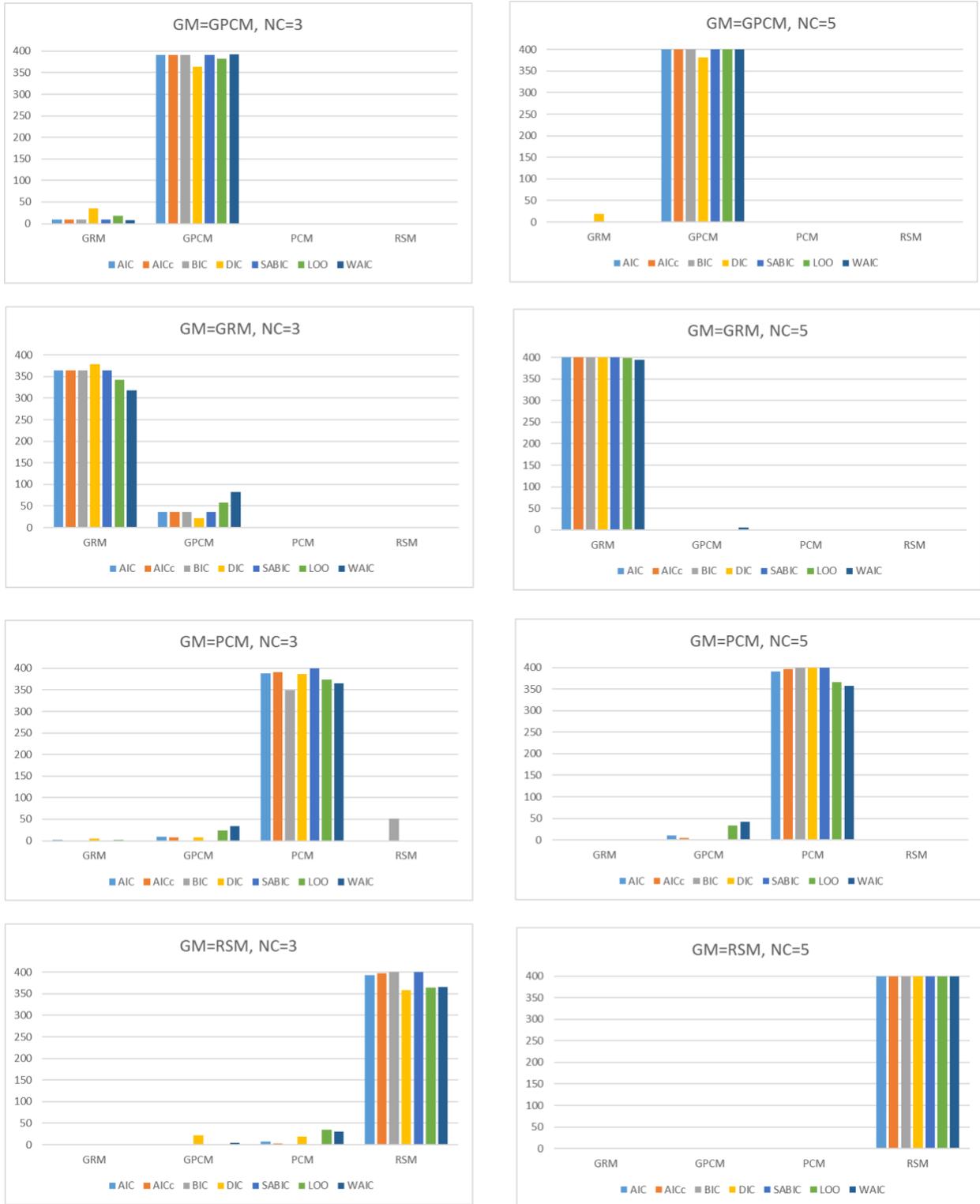

*Figure 4.* Model selection by number of categories

**A Real Data Example**

In this section, we demonstrate with a real data set the use of the seven model selection methods to choose the best fitting polytomous IRT model. Data were extracted from student responses to the Verbal Session of the General Aptitude Test (GAT-V), a high-stakes test used in Saudi Arabia for university admission purposes. GAT-V consists of 52 multiple-choice items that are scored dichotomously, and 20 of them are reading comprehension items. We created four polytomous items by extracting items nested within four reading comprehension passages and summing up the item scores within each passage. Since there are three items within each of the chosen, the created polytomous items have four response categories and the score range for each polytomous item is from zero to three. There are 4,960 examinees in the current data set, and we fit the four polytomous IRT models to the 4,960 by 4 item response matrix and compute the seven model fit indices. It should be noted that due to the large sample size, for MCMC estimation we ran three parallel chains of 600 iterations to ensure model convergence and the computation of DIC, WAIC, and LOO was based on simulated samples of 900 posterior draws.

Table 3

*Model Comparison for Four Polytomous Items*

|        | AIC      | BIC      | AICc     | SABIC    | DIC      | WAIC     | LOO      |
|--------|----------|----------|----------|----------|----------|----------|----------|
| GRM    | 43817.8  | 43921.0  | 43817.9  | 43870.2  | 42131.7  | 42308.8  | 42430.7  |
| GPCM   | 43714.4  | 43817.6  | 43714.5  | 43766.8  | 41971.6  | 42002.3  | 42234.4  |
| PCM    | 44012.6  | 44096.5  | 44012.7  | 44055.2  | 42469.4  | 42685.0  | 42873.9  |
| RSM    | 44708.1  | 44753.3  | 44708.2  | 44731.1  | 43202.3  | 43450.7  | 43630.8  |

Table 3 presents the computed model fit indices for GRM, GPCM, PCM, and RSM. As can be seen, all seven model selection methods consistently point to GPCM as the best fitting model, and GRM has the second smallest model fit index value regardless of the model selection method. The fact that all seven model selection methods agree with each other in the current example is hardly surprising given the large sample size: as shown in the previous section, although these methods have difficulty in differentiating between GPCM and GRM with small sample size, with larger sample size they tend to produce more consistent results. PCM has considerably worse model fit than GPCM, suggesting that the constraint imposed by PCM that all items have the same discrimination power is not supported by the data. RSM has the largest model fit index values, which is expected given that it is the most stringent among the four polytomous IRT models.

## Conclusions and Discussions

WAIC and LOO are two fully Bayesian model selection indices that have been shown to perform better than other common indices such as AIC, BIC, and DIC in the context of dichotomous IRT model selection. In the current study, we investigated whether the superior performances of WAIC and LOO in the dichotomous case could be generalized to the context of polytomous IRT model selection. It was found that while both WAIC and LOO had excellent statistical power (mean power greater than 0.93) across the 32 simulation conditions, their performances were slightly worse than the other five model selection methods investigated in this study, namely AIC, BIC, AICc, SABIC, and DIC.

A closer examination of Figures 2-4 reveals why WAIC and LOO had slightly lower statistical power than the other model selection methods. When data was generated based on the GRM (GM=GRM) and either the test length was relatively short (TL=10), the sample size was relatively small (SS=500), or the number of response categories was relatively small (NC=3), WAIC and LOO were noticeably more likely to choose GPCM as the true model. With the increase of test length (TL=20), sample size (SS=1000), or the number of response categories (NC=5), WAIC and LOO performed similarly with other methods. Another scenario where WAIC and LOO performed worse than the other methods was when data was generated based on PCM (GM=PCM): WAIC and LOO were more likely to choose GPCM, the more parameterized model, as the true model. It is interesting to note that such a tendency to incorrectly choose GPCM over PCM only decreased with the increase of test length, but not with the increase of either sample size or number of response categories.

The hypothesis that WAIC and LOO be superior to other model selection methods in the context of polytomous IRT model selection due to their fully Bayesian nature is not supported by the findings of the current study. The four model selection methods based on the frequentist framework (AIC, BIC, AICc, and SABIC) had higher statistical power than their counterparts based on the Bayesian framework (DIC, LOO, and WAIC), among which DIC performed the best. This observed pattern in the case of polytomous IRT model selection is in contrast with what is observed in the case of dichotomous IRT model selection, where fully Bayesian methods (LOO and WAIC) performed better than the partially Bayesian method (DIC), which in turn performed better than the non-Bayesian methods such as AIC, BIC, and LRT.

As the current study was intended as an extension of Kang, Cohen, and Sung (2009) study, it is of interest to compare the two studies to see whether any inconsistencies in the findings have occurred. Before comparisons are made, it should be noted that the same simulation design and the same set of item parameters used for response data generation were used in both studies. The current study used an extension of the powerful Hamiltonian Monte Carlo (HMC) algorithm implemented in Stan for MCMC estimation, while Kang, Cohen, and Sung used the Gibbs sampler implemented in WinBUGS; and due to the advancement of computing power and the efficiency of HMC algorithm over the Gibbs sampler, we were able to increase the number of replications within each simulation condition from 50 to 100 in the current study. In regard to the findings, what is consistent between the two studies is that the frequentist model selection methods were superior to the Bayesian ones: Kang, Cohen, and Sung found in their study that AIC and BIC performed better than DIC and CVLL, and we found that AIC, AICc, BIC, and SABIC performed better than DIC, LOO, and WAIC. What is inconsistent between the two studies is the specific performances of DIC, which seemed to perform considerably worse in Kang, Cohen, and Sung study. For example, in their study when the data generating model was GRM and the number of response category was 3, DIC had a probability of close to 0.5 to choose GPCM as the true model (p. 511) and such a probability did not decrease much with the increase of number of response categories; in the current study, however, DIC only had a probability of about 0.05 to choose GPCM when NC=3, and with the increase of number of response categories, such a probability decreased to almost zero. As both studies used the same set of item parameters and same simulation design for data generation, we believe the different number of replications within each simulation condition (100 vs 50) cannot possibly cause such drastically differences regarding the performance of DIC.

One possible cause could be model convergence issue caused by the use of different MCMC methods: Kang, Cohen, and Sung (2009) used WinBUGS that implements the Gibbs sampler and ran one chain with 11,000 iterations, 5,000 of which were discarded as burn-in iterations; we used Stan that implements HMC algorithm and ran three chains with each having 400 iterations, half of which were discarded as warm-up iterations. To verify whether model convergence was the reason for the inconsistent performances of DIC in two studies, we fit the GPCM to the 100 simulated datasets under one simulation condition (GM=GRM, TL=20, SS=500, NC=5) using WinBUGS with the same priors and number of iterations as in Kang, Cohen, and Sung (2009), and instead of one Markov chain we ran three to facilitate the check of model convergence. As for each data set there is 620 parameters to be estimated with GPCM (500 person parameters, 20 item discrimination parameters, 20 item location parameters, and 80 item step parameters), overall there are 620,000 parameters estimated across the 100 datasets. Among the 620,000 PSRF values corresponding to the estimated parameters, with WinBUGS estimation there are 322 values greater than 1.1 and 664 values greater than 1.05; in addition, the maximum PSRF value is 6.01. In contrast, with Stan estimation the maximum PSRF value is 1.08, and there are only four values greater than 1.05. In other words, using WinBUGS with 11,000 iterations for the estimation of GPCM under the chosen simulation condition still resulted in some cases where the model did not converge, while using Stan with 400 iterations enabled model convergence across all 100 datasets, a difference which we believe to be the reason why DIC performs so differently in the two studies.

Among the seven model selection methods investigated in this study, LOO and WAIC had the lowest statistical power to detect the true model in the case of polytomous IRT model selection. This is somewhat counterintuitive given their superior performances in the case of dichotomous IRT model selection, which Luo and Al-Harbi (2017) attributed to their being fully Bayesian. Such a fully Bayesian nature does not seem to translate into better performances when it comes to the choice of a polytomous IRT model among several candidates. Despite their having the lowest statistical power among the seven model selection methods, WAIC and LOO are plausible Bayesian model selection methods that can be used for polytomous IRT model selection given the fact that their mean statistical power rates are greater than 0.93 and they can be easily computed through the combination of R packages **rstan** and **loo**; although DIC has slightly higher statistical power, there is not a readily available package that computes DIC based on **rstan** output and the users may have to write their own functions to compute DIC. WinBUGS does allow the computation of DIC, but as shown previously, the Gibbs sampler implemented in WinBUGS is much less efficient than the HMC algorithm adopted in Stan and consequently, it require considerably longer time to run before model convergence can be reached. Therefore, if a Bayesian method is preferred, we recommend the use of LOO and WAIC for polytomous IRT model selection due to their acceptable statistical power and ease of computation. If not, the four methods based on the frequentist framework, especially AICc and SABIC, should be used.

The current paper only focuses on model selection, but it is worth reiterating that model selection and model fit check are two integral and complementary parts of any model checking endeavor. Being able to choose the best fitting model is no guarantee that the chosen model fits the data if there is not a true model among the candidate ones, and model fit check procedures should always be used in tandem with model selection methods.